\documentclass[12pt]{article}
\usepackage{astrobib,graphicx}
\newcommand{\ltapprox}{\raisebox{-0.5ex}{$\,\stackrel{<}{\scriptstyle
\sim}\,$}}
\begin{document}
%
%
\title{The Evolution of Shocks in Blazar Jets}
%


\author{Geoffrey V. Bicknell, $^{1,2}$ 
 Stefan J. Wagner, $^{3}$ 
} 

\date{}
\maketitle

{\center
$^1$ Research School of Astronomy \& Astrophysics, Mt. Stromlo Observatory, Cotter Road, Weston,
ACT 2611\\Geoff.Bicknell@anu.edu.au\\[3mm]
$^2$ Department of Physics \& Theoretical Physics, ANU, Canberra ACT 0200 \\[3mm]
$^3$ Landessternwarte, Koenigstuhl, D-69117 Heidelberg, Germany\\S.Wagner@lsw.uni-heidelberg.de\\[3mm]
}

%
\begin{abstract}
We consider the shock structures that can arise in blazar jets as a consequence of
variations in the jet flow velocity. There are two possible cases: (1) A double shock system consisting
of both a forward and reverse shock (2) A single shock (either forward or reverse) together with a
rarefaction wave. These possibilities depend upon the relative velocity of the two different sections of jet.
Using previously calculated spherical models for estimates of the magnetic field and electron number density of the
emission region in the TeV blazar Markarian 501, we show that this region is in the form of a thin disk, in the
plasma rest frame. It is possible to reconcile spectral and pair opacity constraints for MKN~501 for Doppler factors
in the range of 10--20. This is easiest if the corrections for TeV absorption by the infrared background are not
as large as implied by recent models.
\end{abstract}


\bigskip

\begin{center}
Accepted for publication in the PASA, {\bf 19}, No. 1. {\em AGN Variability across the Electromagnetic Spectrum}.
See \underline{http://www.publish.csiro.au/journals/pasa/pip.cfm}
\end{center}
%
%

\section{Introduction}

The current phenomenology used to estimate parameters in models of blazars often assumes a spherical
homogeneous emitting region (in the plasma rest frame). This approximation is based in part on the ease of
calculating the inverse Compton spectrum in a spherical geometry. This approach has served the subject well and
has produced useful estimates of the parameters of many blazars \cite{ghisellini97a}. However, when we consider the
reconciliation of constraints based upon spectral breaks and pair opacity, spherical models are probably too
contrived. Furthermore, consideration of the way in which shocked regions would arise in jets lead us away from
the idea of spherical emitting regions. In this paper we take the first steps in the direction of determining the
geometry of blazar emission and how this relates to the dynamics of the underlying flow. This indicates the
direction for future blazar models and also indicates how models based upon more realistic geometries may be better
able to constrain jet Doppler factors and the opacity of the diffuse infrared background.

\section{Spherical models for Markarian 501}

\begin{figure}[ht!]
\centering \leavevmode
\includegraphics[width=\textwidth]{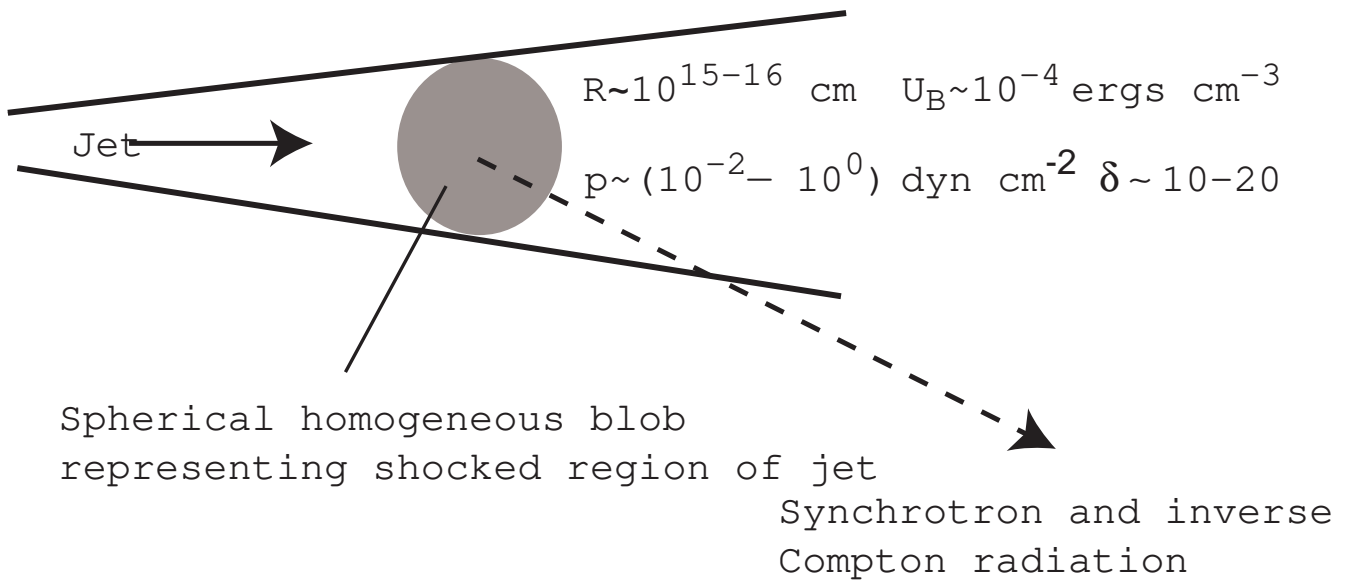}
\caption{A schematic indication of the spherical blob model for the emission region in blazar jets.}
\label{f:spherical_blob}
\end{figure}

Let us begin by noting the results obtained by fitting spherical emission models for the blazar Markarian~501.
The geometry of such models is indicated in Figure~\ref{f:spherical_blob}. Typical parameters for MKN~501, which
emits synchrotron X-rays and inverse Compton TeV
$\gamma$-rays, inferred from this geometry are a radius, $R \sim 10^{15-16} \> \rm cm$, a magnetic field, $B \sim
0.1 \> \rm G$, magnetic energy, $U_B \sim 10^{-4} \> \rm ergs \> cm^{-3}$, particle pressure, $p \sim 10^{-2} - 
10^0 \> \rm dynes \> cm^{-2}$ and a Doppler factor, $\delta \sim 10-20$. 

\begin{figure}[ht!]
\centering \leavevmode
\includegraphics[height=10 cm]{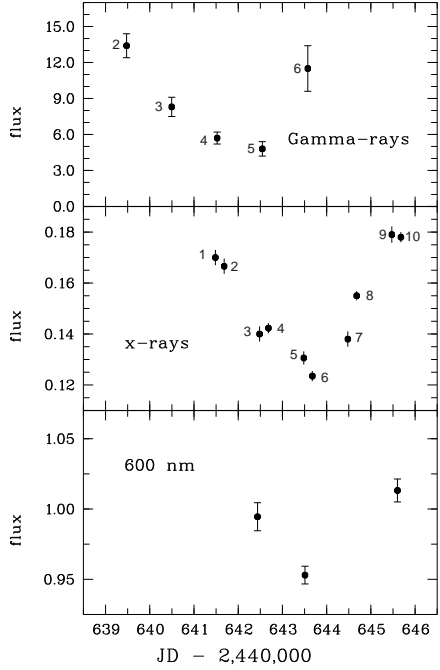}
\caption{The integrated TeV count rate, the RXTE 1 keV PCA count rate and the variation in optical
magnitude for the July, 1997 flare of MKN~501 (Lamer \& Wagner 1998). Various epochs in the TeV and RXTE
observations, referred to in the text, are indicated. The time of onset of the second flare is estimated (mainly
from the RXTE spectra) to be at $t \approx 641.5 \> \rm days$}
\label{f:allflux}
\end{figure}

\begin{figure}[ht!]
\centering \leavevmode
\includegraphics[height=15 cm]{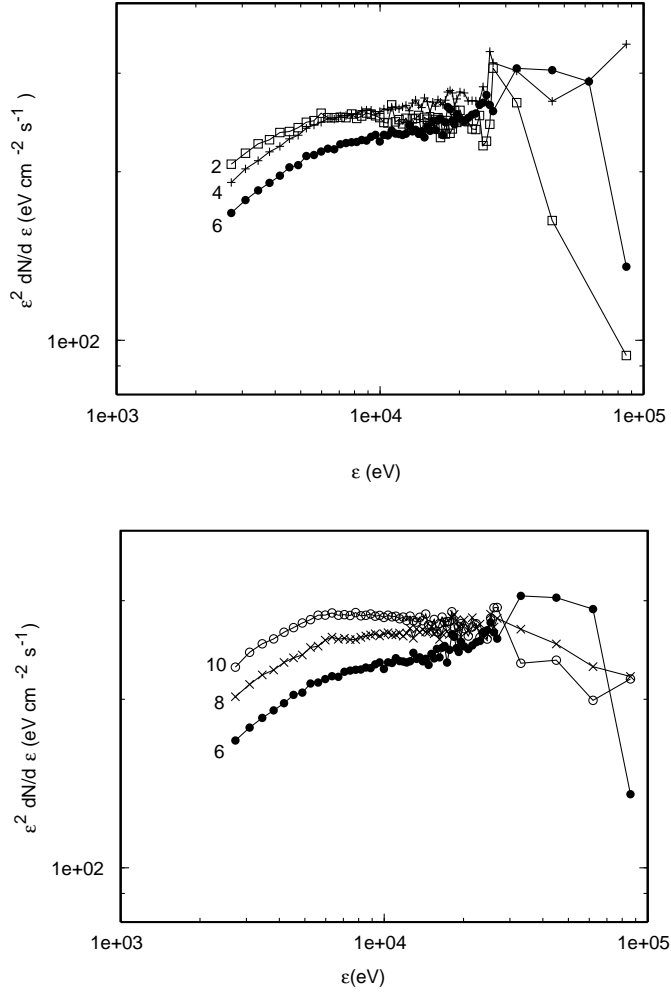}
\caption{The development of the RXTE spectrum (Lamer \& Wagner, 1998) over the period of the X-ray
observations. The even epochs only are shown since these involved longer integration times than the odd
numbered epochs providing adequate signal to noise in the 20--100~keV range. Error bars have been
left off in order to distinguish the different spectra.}
\label{f:epochs2-10}
\end{figure}

MKN~501 flares sporadically in X--rays and
$\gamma$--rays. Figure~\ref{f:allflux} shows the evolution of the TeV $\gamma$--rays, 1 keV RXTE PCA flux and
optical emission from MKN~501 during part of a flare that occurred in July, 1997 \cite{lamer98a}.  The period of
observation encompasses a subsiding flare and a newly developing one, necessitating two component models for the
flare. Note that the 1 keV X-ray flux does not rise until about a day after the $\gamma$--ray flux. The reason for
this is apparent in the time-dependent RXTE spectra \cite{lamer98a} shown in figure~\ref{f:epochs2-10}.
Whilst the low energy part of the spectrum decreases, a high energy part starts to emerge at about
epoch 4 and dominates in epoch 6. By epoch 10, the spectrum has returned to a form similar to that in
epoch~2 exhibiting a break in spectral index of approximately 0.3 at approximately 6~keV.

\begin{figure}[ht!]
\centering \leavevmode
\includegraphics[width=\textwidth]{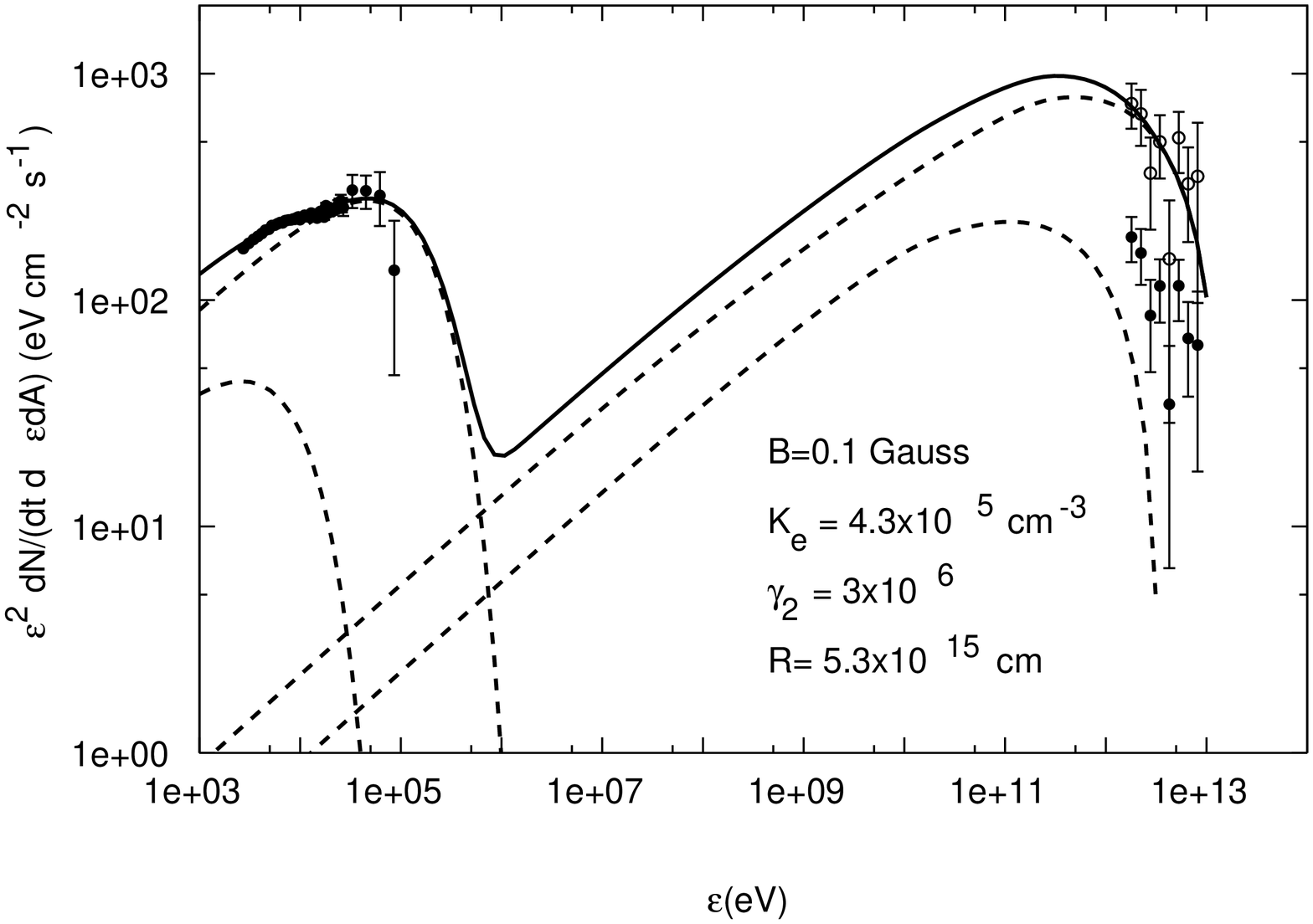}
\caption{A fit to the combined epoch 6 X-ray and epoch 6 TeV $\gamma$--ray data from the July, 1997
flare from Markarian~501 assuming a Doppler factor of 10 (Bicknell, Wagner \& Groves, 2001). The TeV data have been
corrected using the opacity of the IR background derived by Guy et al. (2000). The raw data are shown as solid
circles; the corrected data are shown as open circles. However, the extent of the required correction is currently
uncertain.}
\label{f:501_fit}
\end{figure}

A two component spherical model for the combined epoch 6 X--ray and epoch 6 $\gamma$--ray data\footnote{The
$\gamma$--ray observations did not extend beyond epoch 6 because of increasing moonlight.} is shown in
figure~\ref{f:501_fit}. Two synchrotron components based on energy--truncated power--law spectra were fit
to the X-ray data. We take the point of view that the high energy ($\hbox {break energy} \sim 50-100 \>
\rm keV$) component whose emergence is apparent during earlier epochs is related to the $\gamma$--ray
flare. The low energy component probably encompasses the lower energy X-rays from both the subsiding and
developing flares. In order to restrict the number of parameters we have assumed that the magnetic field
is the same in each component. The model parameters for each component are: Doppler factor, $\delta$, the upper
cutoff in Lorentz factor of the electron distribution, $\gamma_2$, the magnetic field (in the rest frame), $B$, 
the normalising constant of the electron distribution, $K_e$ and the electron spectral index, $a$. (The electron
number density per unit Lorentz factor is $N(\gamma) = K_e \gamma^{-a}$.) The parameters reported in the figure
legend relate only to the high energy component. The parameters of a number of model fits to this epoch of MKN 501
are given in table~\ref{t:mkn501_fits}. These models include examples of where a correction has been made to the
infrared background and examples of where no correction has been made. In relating the bulk Lorentz factor,
$\Gamma$ to the Doppler factor we have assumed that the jet is nearly pole-on, so that $\delta \approx 2 \Gamma$.
The bulk Lorentz factor is not used in the models but is used for other purposes below.

\begin{table}[ht!]
\begin{center}
\begin{tabular}[t]{| c | c | c | c | c | c | c | c | c | c |}
\hline
Model & $\delta$ & $\Gamma$  & $\gamma_2$ & $a$ & $B$    & $K_e$                 & $p_e$                      & 
$u_B$                &
$R$               \\
&         &           &         &   & Gauss     & $\rm cm^{-3}$ & $\rm dynes \; cm^{-2}$ & $\rm ergs\; cm^{-3}$
&
$ 10^{15} \> \rm cm$ \\
\hline 
\multicolumn{10}{|l|}{IR background corrected} \\
\hline
1 & 10       & 5         & $3 \times 10^6$   & 2.0 &  0.1   &  $4.3 \times 10^5$    & 1.2                       &
$4.0
\times 10^{-4}$    & 5.3              \\ 
2 & 20       & 10        & $1.5 \times 10^6$ & 2.0 &0.2    &  $2.6 \times 10^6$         &
6.8                        &
$1.6
\times 10^{-3}$    & 0.87             \\
\hline
\multicolumn{10}{|l|}{Not corrected for IR background} \\
\hline
3 & 10       & 5         &  $3 \times 10^6$  & 2.0 & 0.06  & $4.4 \times 10^3$    & 0.012              &
$1.4
\times 10^{-4}$    & 14              \\ 
4 & 20       & 10        &  $ 1.5 \times 10^6$ & 2.0 & 0.13  & $5.4 \times 10^4$    &   0.14           &
$6.7
\times 10^{-4}$  & 1.5               \\ 
\hline
\end{tabular}
\end{center}
\caption{Comparison of parameters of synchrotron plus inverse Compton model fit to data on MKN~501. The emitting
region is assumed to be spherical in the rest frame.}
\label{t:mkn501_fits}
\end{table}

A typical fit to the MKN~501 flux emitted during a large flare in July, 1997 is shown in Figure~\ref{f:501_fit}.
In this fit, the TeV data have been corrected for absorption by the diffuse infrared background using the opacity
estimated by \citeN{guy00a}. However, the extent of the correction is currently
controversial and we have also summarised in table~\ref{t:mkn501_fits} the model parameters derived when no
correction is made. It is worth emphasizing that the inferred radius of the spherical region is a
parameter that is provided by the model fits. Variability provides an a priori upper limit on this parameter,
namely, that for a variability timescale,
$\Delta t$, the emitting region needs to be small enough that variations are not smeared out. This leads to
\begin{equation}
R < \frac{1}{2} \, c \Delta t \, \delta\ \approx 1.3 \times 10^{16} \, 
\left( \frac {\Delta t}{\rm day} \right) \, \left( \frac{\delta}{10} \right) \> \rm cm
\end{equation}
However, in the spherical model, there is no other {\em a priori} basis for radii
$\sim 10^{15-16} \> \rm cm$. Sometimes the diameter of the sphere is associated with an adiabatic cooling
length. However, this idea has not really been developed significantly.

\section{Production of shocks in jets}

\begin{figure}[ht!]
\centering \leavevmode
\includegraphics[width=\textwidth]{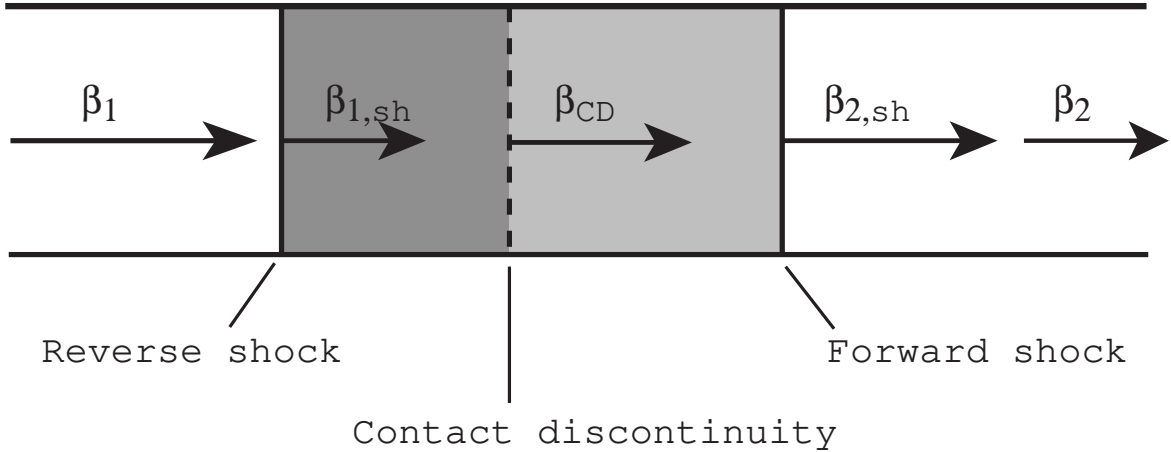}
\caption{The production of forward and reverse shocks in a jet resulting from variations in the flow velocity. Note
that the relative transverse and longitudinal sizes are not to scale. Modelling shows that the longitudinal extent
is generally much smaller than the radial extent. (See table~2).}
\label{f:2shock}
\end{figure}

\begin{figure}[ht!]
\centering \leavevmode
\includegraphics[width=\textwidth]{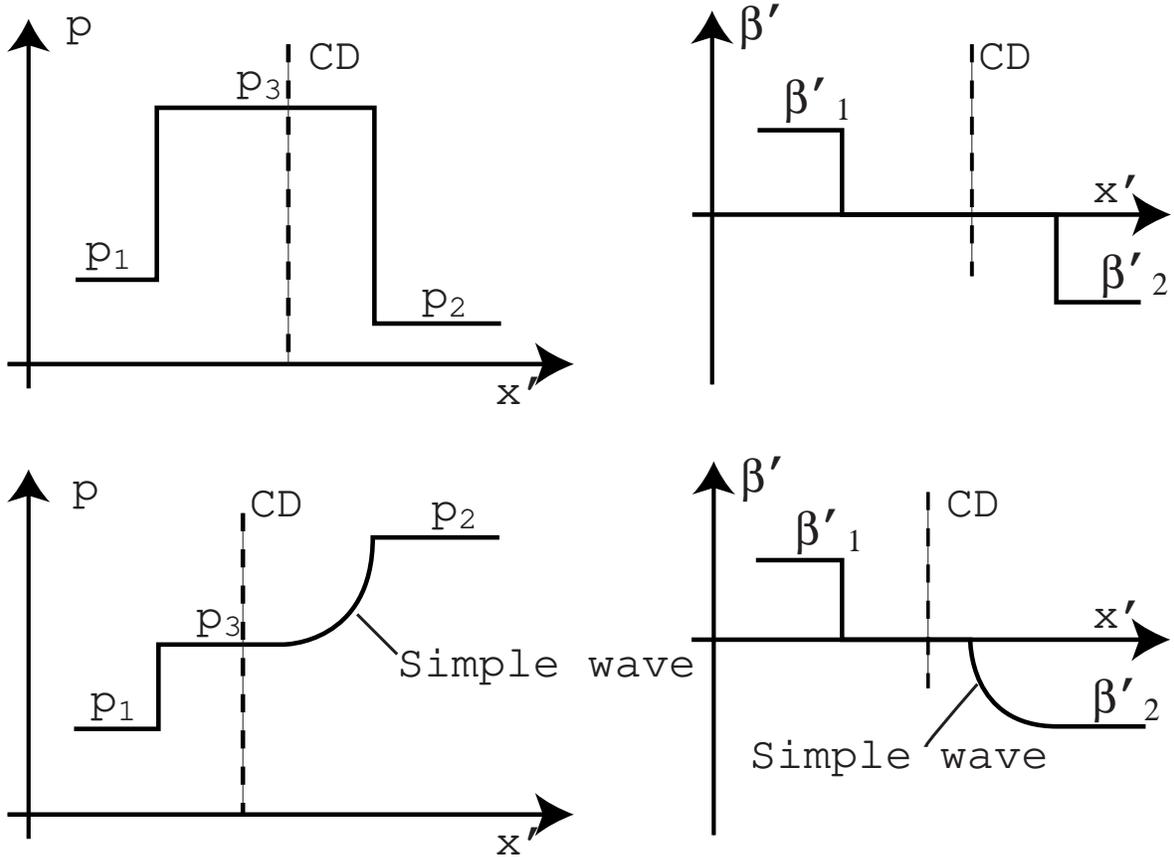}
\caption{The behaviour of pressure and velocity in a relativistic shock tube. The upper part of the diagram
is for the case wherein a forward and reverse shock form; the lower part of the diagram is for the case of a
reverse shock plus a relativistic simple wave. The frame of the contact discontinuity is also the rest frame of
the emitting plasma.}
\label{f:shock_tube}
\end{figure}

In order to make further progress it is important to  understand the possible processes that can lead
to the production of shocks in jets. One process that is often mentioned (e.g. \citeN{rees78a}) is the
production of shocks resulting from variations in the flow velocity. We idealise the
initial conditions as a velocity discontinuity with faster moving gas catching up to
slower moving gas (see figure~\ref{f:2shock}). This is a classic shock tube in which the
equation of state and flow velocities are relativistic. There are two solutions of interest: (1) Two
shock waves are produced, one moving into the gas ahead of the shock -- the forward shock, the other
into the gas behind -- the reverse shock. (2) A single shock (either forward or reverse) and a
rarefaction are produced. The direction of the shock in this case depends upon the details of the
initial parameters. The development of the two possible solutions are sketched in Figure~\ref{f:shock_tube}. 

Both of these situations are most readily analysed by transforming to the frame
of the contact discontinuity separating the two gases. This is the approach used by \citeN{landau87a}
for the non-relativistic case. It is also useful here since, as before, this is the rest frame of the
two gases on either side of the contact discontinuity and is the appropriate frame in which to
calculate the emissivity.

\subsection{Forward and reverse shocks}

Let us first consider the case of a forward and reverse shock. It is useful to calculate the difference
in the velocities of the two shocks given the (uniform) pressure,
$p_3$, of the gas sandwiched in between them. In so doing we use the ultrarelativistic equation of state,
$p = 1/3 \epsilon$ where $p$ is the pressure and $\epsilon$ is the energy density. This is
equivalent to assuming that at this stage of the jet, some 100 gravitational radii from the core,
there is no entrained thermal matter and that the initial matter ejected from the black hole is
ultrarelativistic (e.g. is composed exclusively of relativistic electron-positron pairs). The most
straightforward way to do the calculation is in the frame of the contact discontinuity (CD) separating
the two shocked regions, assign the ratio $p_3/p_1$, and then use this to calculate velocities of both shocks
and the velocities of upstream and downstream gas. 

In the frame of the contact discontinuity, the velocities of upstream and downstream gas in regions 1
and 2 may be derived from the expressions given in \citeN{landau87a} for the relative velocity between
pre-shock and post-shock gases, taking into account that in the CD frame the post-shock gas is
stationary. The result is:
\begin{eqnarray}
\beta_1^\prime &=& \frac {\sqrt 3 (p_3/p_1 -1)}{\sqrt {(3 + p_3/p_1)(1+3p_3/p_1)}} \\
\beta_2^\prime &=& - \frac {(p_3/p_1 - p_2/p_1)}{\sqrt {(3p_2/p_1+p_3/p_1)(p_2/p_1+3p_3/p_1)}}
\end{eqnarray}

At this point we introduce a convenient notation for the relativistic composition of velocities. We
denote the relativistic addition and subtraction of velocities by
\begin{eqnarray}
\beta_1 \oplus \beta_2 {\buildrel \rm def \over =} \frac {\beta_1 + \beta_2}{1 + \beta_1 \beta_2} \\
\hbox{and} \qquad \beta_1 \ominus \beta_2 {\buildrel \rm def \over =} \frac {\beta_1 - \beta_2}{1 -
\beta_1 \beta_2}
\end{eqnarray}
respectively.

The {\em frame-independent} relative velocity between the gas on either side of the forward/reverse
shock structure
\begin{equation}
\beta_{12}  = \beta_1 \ominus \beta_2 =
\beta_1^\prime (p_3/p_1, p_2/p_1) \ominus \beta_2^\prime (p_3/p_1,p_2/p_1) 
\end{equation}
may be used to either calculate the relative velocity in the observer's frame as a function of the
parameters $p_3/p_1$ and $p_2/p_1$ or to numerically determine the ratio $p_3/p_1$ of the shock
to  upstream pressures as a function of the relative velocity and the ratio, $p_2/p_1$ of
the downstream to upstream pressures.

\begin{figure}[ht!]
\centering \leavevmode
\includegraphics[width=\textwidth]{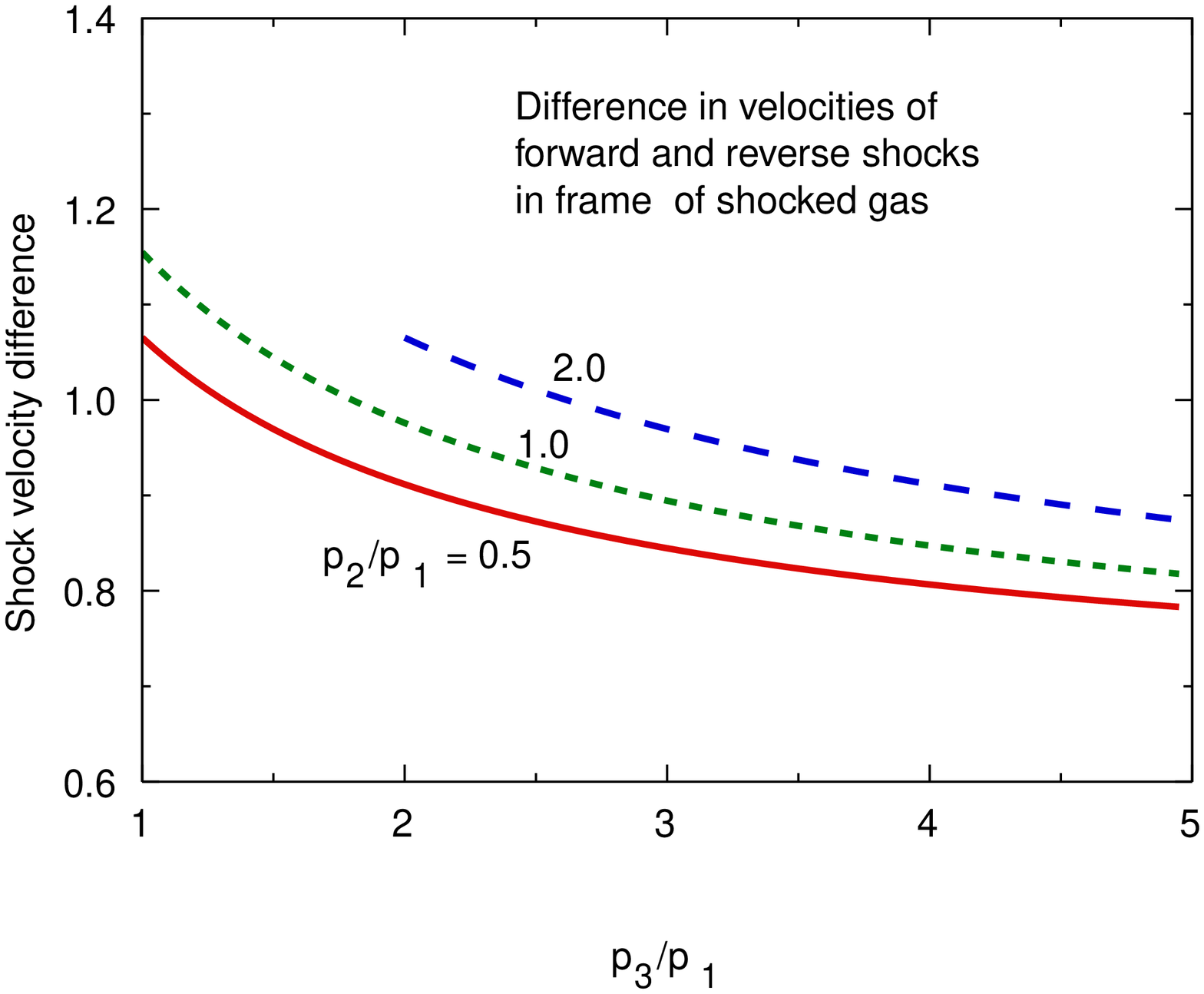}
\caption{The difference in velocities of forward and reverse shocks in the case when two shocks are produced.}
\label{f:vsh_p3}
\end{figure}

What is most important for the current purposes is the difference in the {\em shock} velocities since
this determines the size of the emitting region in the rest frame. These may be calculated from the
expressions given in \citeN{landau87a} for the pre-shock and post-shock velocities {\em in the shock
frame.} Denoting the shock frame by a double prime superscript, then the velocity of the reverse shock,
for example, in the CD frame, is given by:
\begin{equation}
\beta_1^\prime \ominus \beta_{1,\rm sh}^\prime = \beta_1^{\prime\prime} 
\Rightarrow \beta_{1,\rm sh}^\prime = \beta_1^\prime \ominus \beta_1^{\prime\prime}
\end{equation}
The results for both forward and reverse shocks are:
\begin{eqnarray}
\beta_{1,\rm sh}^\prime &=& - \sqrt {\frac {3+p_3/p_1}{3(1+3p_3/p_1)}}\\
\beta_{2,\rm sh}^\prime &=& \sqrt {\frac {3+p_3/p_2}{3(1+3p_3/p_2)} }
\end{eqnarray}
The limits of strong and weak shocks are informative. 
\begin{equation}
\begin{array}[]{l c l l}
\hbox{As} & p_3/p_1 \rightarrow \infty &  
\beta_{1,\rm sh}^\prime, \beta_{2,\rm sh}^\prime \rightarrow \mp 1/3 \\
\hbox{As} & p_3/p_1 \rightarrow  1  & 
\beta_{1,\rm sh}^\prime,  \beta_{2,\rm sh}^\prime \rightarrow \mp 1/\sqrt 3
\end{array}
\end{equation}
These limits are to be expected from the limiting velocities in the post-shock region of a relativistic
flow, remembering that the post-shock region is at rest in this frame. Note also that the range in
velocities between strong and weak shocks is not great -- approximately $0.33- 0.58 \, c$. The
shock velocity difference, which determines the size of the emitting region, is shown in
figure~\ref{f:vsh_p3}. Note that the size of the region is {\em not} determined by the relative
velocity. However, the relative velocity has a role which we consider further below.

The size of the flaring region in the rest frame as a function of the time in the rest frame, $\Delta
t^\prime$ and the observer's time, $\Delta t$, is given by:
\begin{eqnarray} \Delta x_{\rm sh}^\prime &=& 
|\beta_{1,\rm sh}^\prime - \beta_{2, \rm sh}^\prime| \, c \Delta t^\prime = 
\frac {|\beta_{1,\rm sh}^\prime - \beta_{2, \rm sh}^\prime|}{\Gamma_{\rm CD}} \, c \Delta t \\
&\approx& 2.6 \times 10^{14} \> |\beta_{1,\rm sh}^\prime - \beta_{2, \rm sh}^\prime| \,
\left( \frac {\Gamma_{\rm CD}}{10} \right)^{-1} \, \left( \frac {t}{1 \rm day} \right) \> \rm cm
\end{eqnarray}
Time dilation between the rest and observer's frame has been incorporated. Since the difference in
shock velocities does not vary greatly with the pressure of the shocked gas, we adopt a fiducial value 
of unity for that parameter. 

An immediately interesting feature of this result is that for a Lorentz factor of 5 (corresponding to a
Doppler factor $\sim 10$) and timescales of a few days, the size of the shocked region in the direction
of flow is comparable to the radii inferred from fits of the spherical model. This is a direct
consequence of the insensitivity of the size of the shocked region to its pressure.

\subsection{The relative velocity and the condition for two shocks}

We have already seen that the pressure in the shocked region can be estimated from the relative
velocity of the unshocked gas on either side of the reverse-forward shock region. The relative
velocity also enters into the condition for a forward--reverse shock pair to occur. The relative
velocity needs to be large enough to produce a pressure,
$p_3$, between the shocks that is greater than both $p_1$ and $p_2$. The limiting case occurs when $p_3
= {\rm max} (p_1,p_2)$. This leads to the following condition on the relative velocity, $\beta_{12}$ of
the gas on either side of the shocked region:
\begin{equation}
\beta_{12} > \sqrt 3 \frac {|p_2/p_1 -1 |}{\sqrt{(3 p_2/p_1+1)(3 + p_2/p_1)}} 
\label{e:b12}
\end{equation}

\begin{figure}[ht!]
\centering \leavevmode
\includegraphics[width=\textwidth]{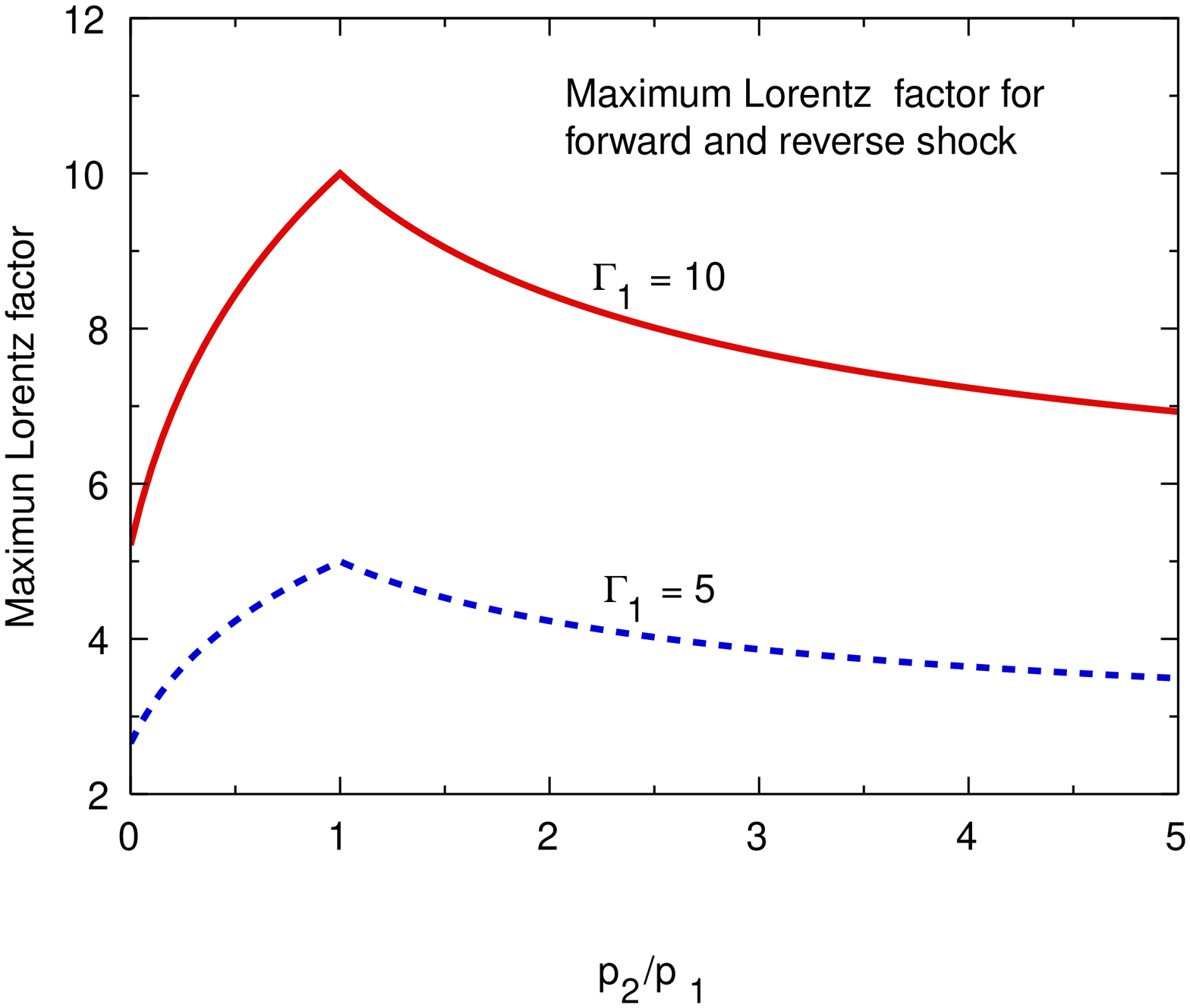}
\caption{This plot shows the maximum Lorentz factor of the gas being overtaken if both a forward and reverse shock
are to result. The maximal Lorentz factors are shown for two values, 5 and 10, of the Lorentz factor of the
overtaking gas.  }
\label{f:lf_max}
\end{figure}

If the relative velocity is less than this critical value, that is, if the variation in velocity is not
substantial enough, then a forward--reverse shock structure does not form. In order to gain some
numerical insight, one can assign a Lorentz factor to the overtaking gas (region 1) and use
equation~(\ref{e:b12}) to calculate the {\em maximum} Lorentz factor of the gas ahead of
the forward shock (region 2) for a two-shock solution to occur. The maximum lorentz factor is shown in
figure~\ref{f:lf_max} for overtaking Lorentz factors, $\Gamma_1$ of 5 and 10. 
The relevant parameter
here is the ratio of the pressures downstream of the forward shock, $p_2$ and upstream of the reverse
shock, $p_1$. For example, if 
$p_2/p_1=2$ and $\Gamma_1 = 5$ then the maximum Lorentz factor of the upstream gas is about 4.3. If the
Lorentz factor of the upstream gas exceeds this value then an intermediate shocked region with pressure
greater than both $p_2$ and $p_1$ cannot be supported. The outcome is a rarefaction wave moving into
the downstream gas and a reverse shock moving into the upstream gas. When $p_1 > p_2$ the situation is
reversed.
 
As in the non-relativistic case (see \citeN{landau87a}), the rarefaction region is described by a
centred, simple wave \cite{liang77a}. Simple waves are based on the Riemann invariants, $J_\pm$, of the
one-dimensional gas dynamics equations and for the equation of state $p = \epsilon/3$
\begin{equation}
J_{\pm} = \frac {1}{2} \ln \frac {1+\beta}{1-\beta} \pm \frac {\sqrt 3}{4} \ln p
\end{equation}
In the region of right propagating simple wave, $J_- = \hbox{constant}$ and the equations of the right
propagating characteristics are:
\begin{equation}
\frac {x}{ct} = \frac {\beta + 1 / \sqrt 3}{1- \beta / \sqrt 3}
\end{equation}
These two features of a centred simple wave allow one to develop the solution in the rarefaction region
and patch it to the solution on the left representing the reverse shock (see figure~\ref{f:shock_tube}). As
with the two-shock solution it is convenient to develop the shock plus rarefaction solution in the
frame of the contact discontinuity and in terms of the pressure, $p_3$ in the post-shock region. In this
frame, the unshocked gas velocity on the left is given by:
\begin{equation}
\beta_1^\prime = \frac {\sqrt 3 (p_3/p_1 -1)}{\sqrt{(3 + p_3/p_1)(1 + 3 p_3/p_1)}}
\end{equation}
and the velocity on the right is given by:
\begin{equation}
\beta_2^\prime = \frac {(p_2/p_3)^{\sqrt 3 /2}-1}{(p_2/p_3)^{\sqrt 3 /2}+1} 
\end{equation}
The frame-independent relative velocity,  $\beta_{12} = \beta_1^\prime \ominus \beta_2^\prime$ can be
used to solve numerically for the pressure, $p_3$ in terms of $\beta_{12}$ if required.

In this solution, the velocity of the reverse shock in the contact discontinuity frame (which is also
the rest frame of the shocked gas) is the same as that of the reverse shock in the two-shock case, i.e.
\begin{equation}
\beta_1^\prime = - \sqrt {\frac {3 + p_3/p_1}{3 (3p_3/p_1+1)}}
\end{equation}
and the size of the shocked region in the CD frame as a function of the observer's time is also similar,
viz.:
\begin{equation}
\Delta x^\prime = | \beta_1^\prime | \frac {c \Delta t}{\Gamma} 
\approx 2.6 \times 10^{14} | \beta_1^\prime | \, \left( \frac {\Gamma}{10} \right)^{-1} \,
\left( \frac {t}{\rm day} \right) \> \rm cm
\end{equation}

\section{Application to the flare in Markarian 501 -- the size of the shocked region}

Inspection of the RXTE spectra and the $\gamma$--ray light curve show that the second flare begins at
$t \approx 641.5 \> \rm days$ where $t$ refers to JD--2,440,000. This means that epoch~6 for which we have
presented the two component models (see figure~\ref{f:501_fit}) corresponds to $\Delta t \approx 643.7 - 641.5  =
2.2 \> \rm days$ and X--ray epoch 10 corresponds to $\Delta t \approx 645.7 -641.5 = 4.2 \> \rm days$.
The entire period surrounding this flare involved such a marked departure from the quiescent flux of Markarian~501 
that it is reasonable to assume that the relative velocity exceeded the critical value derived
above and that
$|\beta_{1,\rm sh} - \beta_{2,\rm sh}| \sim 1$ giving a longitudinal size for the flaring region $\sim
1.1 \times 10^{15} \, (\Gamma/5)^{-1} \> \rm cm$. We neglect for now the complications resulting from calculating
the inverse Compton emissivity in a non-spherical region and, in order to obtain indicative estimates of the
radius of the emitting disk--like region of the jet, we  assume that the volume of the emitting region is the same
as that of the corresponding spherical region as modelled previously. As recorded in table~\ref{t:disk_fits}, this
gives radii ranging from about $9 \times 10^{14} \> \rm cm$ to $4 \times 10^{16}$ cm.  Thus, the emitting region is
a reasonably thin slice of the jet, and in the rest frame, the ratio of its longitudinal extent to its diameter
ranges from 0.01 to 0.3, depending upon the details of the model. The estimate of the longitudinal size arises
naturally from the dynamics of the shocked region and one does not have to appeal to adiabatic expansion to limit
the effective size. This is also the case for the shock plus rarefaction solution. 

\section{Constraints from spectral breaks}

As the flare evolves, by epoch 10 the X--ray spectrum  (see Figure~\ref{f:epochs2-10}) has settled down to a
shape similar to that of the original flare in epoch 2.  We regard this as indicating the establishment of a
quasi--steady state spectrum in which cooling following a shock (or shocks) has established a broken power--law
with the break energy at about 6~keV. This spectral feature provides further constraints on the model with
interesting implications.

The criterion for the location of the break is somewhat contrived in the framework of a spherical model. This
is especially the case with temporal data since there is really no criterion for how the radius of the
emitting region would evolve. We therefore shall not burden the reader with calculations appropriate for a
spherical geometry and we only consider the implications of the spectral break for the disk model whose dynamics we
have considered above.

A break in spectral index in the context of
shock-generated spectra occurs at that photon energy where the cooling time of electrons is equal to
the travel time of the plasma across the emitting region. In this case this means that the cooling time at the
break energy equals the duration of the flare. Often, only the effect of synchrotron cooling is considered.
However, in many blazars the inverse Compton power is comparable to or greater than the synchrotron power and the
contribution of inverse Compton cooling to the total cooling timescale must be considered.  When inverse Compton
scattering in the Klein-Nishina limit is involved (as it is for TeV blazars) the calculation of the cooling time
is more complicated than it is when the Thomson limit applies. However, in the Klein-Nishina limit, both the
cross-section and energy transfer decrease and we therefore assume that the dominant inverse Compton contribution
occurs for interactions between photons and the radiation field that occur in the Thomson limit. Thus, in
estimating the appropriate radiation energy density for an electron of Lorentz factor, $\gamma$, in the plasma
rest frame, one determines a cutoff photon energy, $\epsilon_2$, in the observer's frame, defined by 
\begin{equation}
\epsilon_2(\gamma) \approx \delta \frac{m_e c^2}{\gamma} \approx 5.1 \left( \frac {\delta}{10} \right)
\left( \frac {\gamma}{10^6} \right)^{-1} \> \rm eV
\end{equation}
For a cylindrical uniformly emitting disk of length, $\Delta l$, and radius, $R$, the radiation energy density
per unit frequency at the centre of the disk in the plasma rest frame and the observed flux density are given
by:
\begin{eqnarray}
u_\nu^\prime &=& \frac {4 \pi j^\prime_{\nu^\prime} }{c} f(\Theta_c)
\\ F_\nu &=& \frac {\delta^3}{D_L^2} \> \pi R^2 \Delta l j^\prime_{\nu^\prime} 
\end{eqnarray}
where
\begin{eqnarray}
\Theta_c &=& \tan^{-1} (\Delta l / 2R)\\
\hbox{and} \qquad f(\Theta_c) &=& \Theta_c - \tan \Theta_c \ln \sin \Theta_c
\end{eqnarray}
For a thin disk, $\Theta_c  \approx \Delta l / 2R $ and $f(\Theta_c) \approx (\Delta l /2R) (1+\ln (2R/\Delta l))$
. Thus the radiation energy density is defined in terms of the flux density, $F_\epsilon = F_{\rm \epsilon_0}
(\epsilon_2/\epsilon_0)^{-\alpha} \> \rm ergs \> cm^{-2} \> s^{-1} \> erg^{-1}$ by:
\begin{equation}
u_{\rm rad}^\prime \approx \frac {4 D_L^2}{cR \Delta l} \, f(\Theta_c) \,
\delta^{-4} \left( \frac {\epsilon_0 F_{\epsilon_0}}{1-\alpha} \right) \, 
\left( \frac {\epsilon_2}{\epsilon_0} \right)^{1-\alpha}
\end{equation}
where $\epsilon_2(\gamma)$ is the cutoff in photon energy (observer's frame) defined above and $\epsilon_0 = 1 \>
\rm keV$ is a convenient fiducial energy.

The synchrotron plus inverse Compton cooling time, corresponding to the break energy, $\epsilon_b$, is,
in the plasma rest frame:
\begin{eqnarray}
\Delta t_c^\prime &=& 2^{1/2} 3^{3/2} \pi  \frac {(m_e c e \hbar)^{1/2}}{\sigma_T} B^{-3/2} 
\, \left[1 + \frac {u_{\rm rad}}{u_B} \right]^{-1} {\epsilon_b^\prime}^{-1/2}  \\
&\approx& 3.23 \times 10^3 \, B^{-3/2} \left[1 + \frac {u_{\rm rad}^\prime}{u_B} \right]^{-1} \, 
\delta^{1/2} \,
\left( \frac {\epsilon_b}{\rm keV} \right)^{-1/2} \> \rm sec
\end{eqnarray}
where $u_B = B^2 /8 \pi$. In order to compare with the observed flare duration, we use the 
corresponding time in the observer's frame:
\begin{equation}
\Delta t_c = \Gamma t_c^\prime \approx 3.23 \times 10^3 \, B^{-3/2} \left[1 + 
\frac {u_{\rm rad}}{u_B} \right]^{-1} \,  \left(\delta \Gamma^2 \right)^{1/2} \,
\left( \frac {\epsilon_b}{\rm keV} \right)^{-1/2} \> \rm sec
\end{equation}

Normally, one would invert the expression for $\Delta t_c$  to obtain an expression for the break energy
$\epsilon_b$ as a function of time. However, $\Delta t_c$ depends both explicitly on the break
energy and implicitly through the dependence of
$u_{\rm rad}$ upon $\epsilon_b$. Therefore, in assessing the implications of the break in the X-ray spectrum, it
is easiest to compare the cooling time, $\Delta t_c$ at the observed break energy with the time of duration of the
flare (approximately 4.2 days at epoch 10). These comparisons are recorded in table~\ref{t:disk_fits} for the
various models summarised in table~\ref{t:mkn501_fits}.

\begin{table}
\begin{center}
\begin{tabular}[t]{| c | c | c | c | c | c | c | c | c | }
\hline
Model & $\Delta l^\prime$       & $\Delta l^\prime$       & Radius           &  $u_{\rm rad}^\prime/u_B$ & $\Delta
t_c$  & 
$\Delta t_c /\Delta t$ &  $\tau_{\gamma\gamma}$ & $\tau_{\gamma\gamma}$  \\
      & Epoch 6          & Epoch 10         & of Disk          &                           &  Epoch 10     &  
Epoch 10 & Diameter  & $\Delta l^\prime$ \\
      & $10^{15} \rm cm$ & $10^{15} \rm cm$ & $10^{15} \rm cm$ &           &          days                  & 
& 10 TeV & 10 TeV \\
\hline
1     & 1.1              &  2.2             & 9.5              &  4.9  &  1.3   & 0.31 & 3.3 & 0.20  \\ 
2     & 0.57             &  1.1             & 0.90             &  8.5  &  0.80  & 0.19 & 1.2 & 0.39  \\
3     & 1.1              &  1.1             & 4.1              &  0.95 &  8.4   & 2.0  & 1.1 & 0.015 \\
4     & 0.57             &  2.2             & 2.0              &  5.3  &  2.3   & 0.55 & 0.15& 0.11  \\
\hline
\end{tabular}
\end{center}
\caption{Indicative parameters for the disk model based upon the approximations described in the text.}
\label{t:disk_fits}
\end{table}

\section{Pair opacity}
Pair opacity provides another important constraint on models of blazars -- particularly TeV blazars. 
TeV photons interact with infrared photons to produce pairs, the rate of production again depending upon the
radiation energy density in the plasma rest frame.  Here we are concerned with the infrared photons internal
to the emitting region. For a power-law photon spectrum of spectral index $\alpha$ extending to at least the
infrared, the absorption per unit length, $a_{\gamma\gamma} \> \rm cm^{-1}$, of a $\gamma$--ray of energy
$\epsilon_\gamma$ can be expressed in terms of the observed flux density, $F_{\epsilon_0}$, and observed
$\gamma$--ray energy, $\epsilon_\gamma$, using results presented in \citeN{svensson87a}, as follows:
\begin{equation}
a_{\gamma\gamma} = \frac {4 \sigma_T}{c}  \eta(\alpha) \frac {d_L^2}{\Delta l R} f(\Theta_c) 
\delta^{-(2\alpha+3)} F_{\epsilon_0} \left[ \frac {\epsilon_0 \epsilon_\gamma}{(m_e c^2)^2} \right]^\alpha
\> \rm cm^{-1}
\end{equation}
An analytical expression for the parameter $\eta(\alpha) \sim 1$ has been calculated by \citeN{svensson87a}.  The
opacity depends upon the Doppler factor to a large power, so that a large Doppler factor can also
render the observation of TeV $\gamma$-rays consistent with the photon energy densities required to produce them.

At first sight it appears that this disk model for the shocked region offers a better prospect for the escape of
$\gamma$--rays because of the reduced path length in the direction of the jet. However, aberration implies that
in the rest frame, photon trajectories are almost perpendicular to the jet thereby increasing the path
length. Specifically, if $\Delta l^\prime$ is the size, in the rest frame, of the disk, if $\theta$ is
the angle between the jet and the line of sight in the observer's frame, and if $c \beta$ is the jet velocity,
then the path length of a ray through the disk originating from the back is given by:
\begin{equation}
\frac{\Delta s}{\Delta l^\prime} = \frac {1-\beta cos\theta}{cos \theta - \beta} 
\end{equation}
(with the proviso that $\Delta L$ has a maximum $\sim $ the diameter of the shocked region.) The value of 
$\Delta s / \Delta l^\prime$ tends to compensate for the smaller value of $\Delta l^\prime$ implied by our disk
model so that one can only expect a modest reduction in $\gamma$--ray opacity compared to a spherical model.
Therefore in table~\ref{t:disk_fits} we have given the $\gamma$--ray opacity for 10~TeV photons
based on the diameter of the emitting disk. The emitting region is only marginally optically thin in this case.

On the other hand, the specific {\em normal} shock configuration, upon which we have based our models represents
the worst case for $\gamma$--ray opacity. If the shock(s) defining the disk, were oblique then
the photons emitted perpendicular to the jet direction in the rest frame would not have such a large path length
and a reduction in opacity by an order of magnitude would result. We have therefore, as an indication of the
reduction in opacity related to this model, have tabulated the optical depth based upon the longitudinal extent
of the slab. Naturally this is much less and indicates the importance of generalising this model to the case of
oblique shocks.

Despite these potential reductions in $\gamma$--ray opacity, it appears that the emission of $\gamma$--rays
originates not far above the $\tau_{\gamma\gamma}=1$ surface -- the gammasphere
\cite{blandford94a,blandford95a}.It is possible therefore, that X-ray flares generated at smaller radii will
have no TeV counterpart. An examination of the combined TeV and X-ray light curves would be interesting in this
regard.

\section{Discussion}

\subsection{Summary}

We have presented a simple theoretical approach to determining the geometry of synchrotron and inverse
Compton emitting regions in blazar jets that is intimately related to the dynamics of the underlying flow. Of
course, other treatments of blazar emission regions that postulate relativistic shocks from the outset (e.g.
\citeN{mastichiadis01a}) would give similar estimates for the size of the emitting region. What we have shown here
is how the emission region patches onto the rest of the jet and how it is determined by the flow parameters of
different sections. For high enough relative velocities, the emission region is bounded by two shocks; for relative
velocities lower than critical the emission region patches on via a simple wave. In the latter case both forward
and reverse shocks are possible depending upon the relative strengths of the pressure in the two regions.

Using this approach, we have shown that the extent (in the jet direction) of the shocked emission region is
similar to, but smaller than, what has previously been inferred from spherical models. Making the
approximation that the inverse Compton emission from a shocked section of jet is the same as from the equivalent
sphere, we have shown how spectral breaks and pair optical depths may be estimated. There are some interesting
differences here between models calculated using TeV spectra that have been corrected to allow for the opacity of
the infrared background (IRB) and comparable models for which no corrections have been applied. Referring to
table~\ref{t:disk_fits}, one can see that the IRB--corrected models are more compact and the estimated cooling
times are less than the observed duration of the flare. Moreover, the pair opacity at 10 TeV is quite significant
even in the most favourable case of basing the opacity on the length of the shocked region. In the case of no IRB
correction, it is feasible that the cooling time and the flare time become equal for Doppler factors between 10
and 20. In this regime also, the pair opacity can be reasonably low. Thus it is possible that the IRB opacity
estimated by
\citeN{guy00a} may be too high. However, in view of the various approximations used, these conclusions can only be
regarded as tentative. However, they do point the way to more detailed models in which the various effects can be
modelled in more detail. Such models should include detailed calculations of inverse Compton emission in
anisotropic radiation fields and the calculation of the post-shock flow, taking into account both particle
acceleration and relativistic fluid dynamics.

\subsection{Relationship to numerical simulations of relativistic jets}

Our approach to modelling shocks in relativistic jets that we have described above is analytic and one can ask if
there is there any point in such a treatment given the impressive simulations of relativistic jets that have been
carried in recent years? (For an introduction to the literature on this work one may consult the recent papers of
\citeN{agudo01a}, \citeN{rosen99a} and \citeN{komissarov97a}.) The answer is yes for several reasons: 
\begin{enumerate}
\item Analytic/semi-analytic approximations of jet dynamics are useful when modelling blazar emission and for
other analytic estimates of jet behaviour. Simple models (provided they capture the relevant physics) streamline
the process of model fitting.
\item The numerical simulations are extremely useful for the purpose for which they were intended,
e.g. in comparing relativistic and nonrelativistic jet stability \shortciteN{rosen99a} or in a qualitative
assessment of the behaviour of superluminal components \cite{agudo01a}. However, for computational reasons related
to the Courant criterion, the jet densities of the simulations do not approach the low values that we expect of
relativistic jets in AGN. To be specific, consider a jet in which the relativistic electron distribution has a
lower cutoff at a Lorentz factor of $\gamma_{\rm min}$; let the pressures of the jet and interstellar medium (ISM)
be $p_{\rm jet}$ and $p_{\rm ISM}$ respectively, and let
$T_{\rm ISM} = 10^7 T_7$ be the temperature of the ISM; let $a$ be the electron spectral index. Then the
ratio of jet electron number density to ISM electron number density is given by:
\begin{equation}
\frac {n_{\rm e,jet}}{n_{\rm e,ISM}} \approx 6 \left( \frac {a-2}{a-1} \right) \, \gamma_{\rm min}^{-1} \,
\left( \frac {p_{\rm jet}}{p_{\rm ISM}} \right) \, \left( \frac {kT}{m_ec^2} \right)
\approx 1.6 \times 10^{-3} \gamma_{\rm min}^{-1} T_7
\end{equation}
for a jet in pressure equilibrium with its surroundings and $a=2.2$. Therefore, for an electron-proton jet and an
ISM temperature $\sim 10^7$, the ratio of densities is of order $10^{-5}$ since $\gamma_{min} \sim 100$
\cite{bicknell01b}. For an  electron-positron jet, $\gamma_{\rm min} \sim 10-100$ and the ratio of mass densities
$ \sim 10^{-8}$. The only simulations that begin to approach these remarkable densities are those of
\citeN{agudo01a}, in which the ratio of mass densities is $10^{-3}$. Of course one can learn a lot from
simulations that depend mainly upon the ordering of densities rather than the absolute values. However, such
simulations do not give a good idea of the interaction of the jet with the external medium since this strongly
depends upon the jet density ratio.
\item It is feasible that one can utilise numerical simulations to understand internal jet dynamics in a jet in
which the internal energy and the mass density are dominated by relativistic particles if the parameter
\begin{equation}
\chi = \frac {\rho c^2}{4 p} \ltapprox 1
\end{equation} 
where $\rho$ is the rest mass density \cite{bicknell94a,bicknell95a}. For the low density simulations of
\shortciteN{agudo01a} 
\begin{equation}
\chi = \frac {\rho_{\rm jet} c^2}{4 p_{\rm jet}} = \frac {\rho_{\rm ISM} c^2}{4 p_{ISM}} \, 
\left( \frac {\rho_{\rm jet}}{\rho_{\rm ISM}} \right) = \frac {\mu m_p c^2}{4 k T_{\rm ISM}} \left( \frac
{\rho_{\rm jet}}{\rho_{\rm ISM}} \right) 
\approx 1.6 \times 10^5 \, T_7^{-1} \left( \frac {\rho_{\rm jet}}{\rho_{\rm ISM}} \right)
\end{equation}
For the parameters of the \shortciteN{agudo01a} simulation, $\chi \sim100$. For one thing, this means that the 
velocities of internal shocks are not relevant to blazar emission regions in such a model. 
\end{enumerate}
One could put several counter-arguments to the above. For example, the interstellar medim temperature could be
higher. But then one would have to reconcile this with the lack of observed high temperature bremsstrahlung from
AGN. In summary, the current simulations are impressive and the various groups involved have boldly moved into a
previously uncharted region of parameter space. However, the simulations do not currently address the physics of
blazars. Clearly that is a challenge for the future. 

\section{Acknowledgments}

We thank the organisers of this workshop for a stimulating meeting. SJW wishes to thank the organisers and the
Heidelberg Sonderforschungsbereich for travel support. We acknowledge constructive criticism of the referee that
contributed to an improved version of this paper.

\section{References}


\end{document}